%% file: hades.tex
\documentclass[10pt,twocolumn]{article} 

\usepackage[utf8]{inputenc}
\usepackage{amsmath}
\usepackage{graphicx}
\usepackage{subcaption}
\usepackage{multirow}
\usepackage{paralist}
\usepackage{tablefootnote}
\usepackage{longtable}
\usepackage[T1]{fontenc}
\usepackage{enumitem}
\usepackage{tikz}
\usepackage{filecontents}
\usepackage{xspace}
\usepackage{ulem}
\usepackage[ruled,vlined,linesnumbered]{algorithm2e}
\usepackage{color}
\usepackage{soul}
\usepackage{floatrow}
\usepackage{booktabs}
\usepackage{xcolor}
\usepackage{balance} 
\usepackage[top=1in, bottom=1in, left=0.75in, right=0.75in]{geometry} 
\usepackage{setspace} 
\usepackage{booktabs} 
\usepackage{makecell} 
\usepackage{amsmath}  
\usepackage[breaklinks]{hyperref} 

\newcommand{\Name}{\textsc{Hades}\xspace}

\newif\ifcolormode
\colormodetrue  

\newcommand{\coloredtext}[2]{%
  \ifcolormode
    \textcolor{#1}{#2}%
  \else
    #2%
  \fi
}

\title{%
  \textbf{\LARGE \Name{}: Hierarchical Adaptable Decoding for \\[0.3em]
  Efficient and Elastic vRAN} 
}

\author{
  Jincao Zhu\textsuperscript{1} \quad
  Kobus Van Der Merwe\textsuperscript{1} \quad
  Xenofon Foukas\textsuperscript{2} \quad
  Bozidar Radunovic\textsuperscript{2} \\
  \small{\textsuperscript{1}University of Utah, \textsuperscript{2}Microsoft} \\
  \small{\texttt{jczhu@cs.utah.edu}}
}

\date{}

\begin{document}
\colormodefalse

\maketitle

\begin{abstract}
\input{abstract}
\end{abstract}

\input{introduction}

\input{relatedworks}
\input{background}

\input{design}

\input{evaluation}

\input{conclusion}

\bibliographystyle{plain}
\bibliography{references}
\balance

\end{document}

%% file: abstract.tex

In cellular networks, virtualized Radio Access Networks (vRANs) enable replacing traditional specialized hardware at cell sites with software running on commodity servers distributed across edge and remote clouds. However, some vRAN functions (e.g., forward error correction (FEC) decoding) require excessive edge compute resources due to their intensive computational demands and inefficiencies caused by workload fluctuations.
This high demand for computational power significantly drives up the costs associated with edge computing, posing a major challenge for deploying 5G/6G vRAN solutions.

To address this challenge, we propose \Name{}, a hierarchical architecture for vRAN that enables the distribution of uplink FEC decoding processing across edge and remote clouds. \Name{} refactors the vRAN stack and introduces mechanisms that allow controlling and managing the workload over these hierarchical cloud resources.
More specifically, \Name{} splits the traditional non-stop run-to-completion iterative FEC decoding process into latency-critical early decoding iterations, i.e., related to MAC processing and early pre-parsing for content identification, and completion decoding iterations, i.e., decoding tasks with larger decoding delay budgets for final data bits extraction. This partitioning provides \Name{} the flexibility to utilize the available midhaul (MH) network for offloading the latency tolerant part of decoding to remote cloud instances, while performing time-sensitive decoding at the edge cloud locations for low-delay processing.
\Name{} controls decoding load distribution between the edge and remote clouds, based on the edge decoding capacity and the offload network bandwidth, thus improving the utilization of edge compute. 

Our results show that \Name{} can provide full decoding capacity with significantly less edge decoding resources by leveraging the existing MH bandwidth for offloading.
In addition, \Name{} provides smooth degradation of the required compute resources when the edge decoding capacity is reduced or underprovisioned. 

%% file: introduction.tex
\section{Introduction}

\begin{figure}
  \centering
        \includegraphics[width=0.9\linewidth]{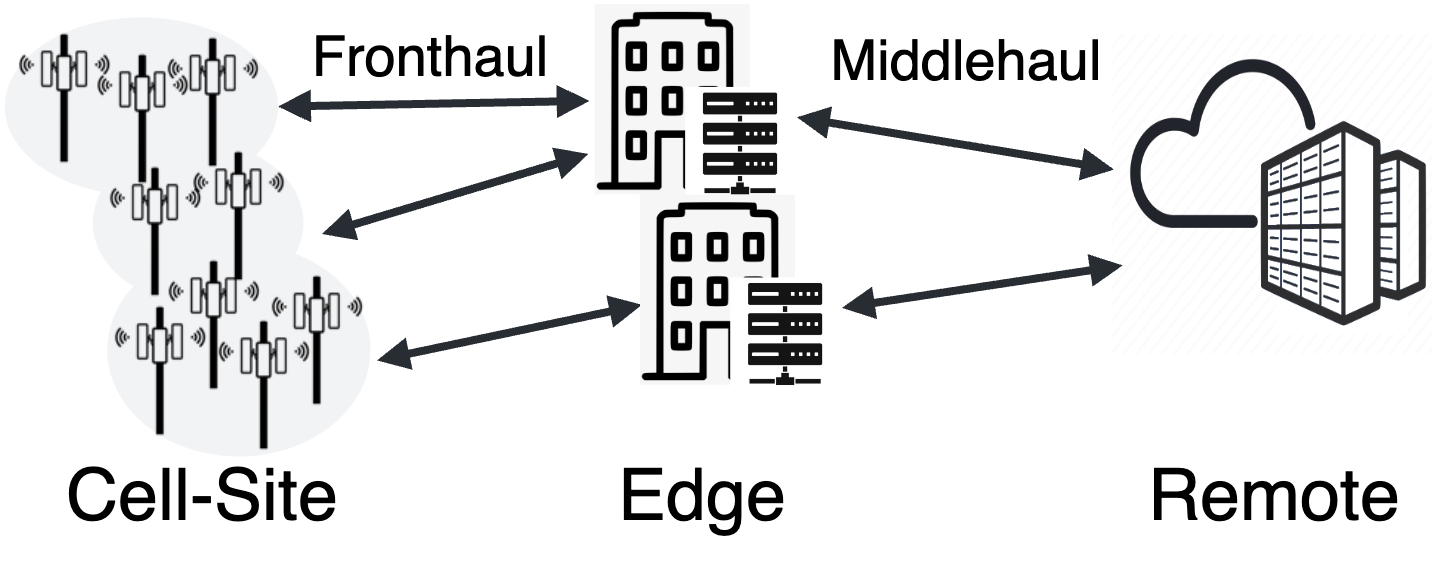}
        \caption{Cell tower, Edge, Remote topology}
        \label{fig:disttopo}
\end{figure}

\begin{figure}
  \centering
  \includegraphics[width=\linewidth]{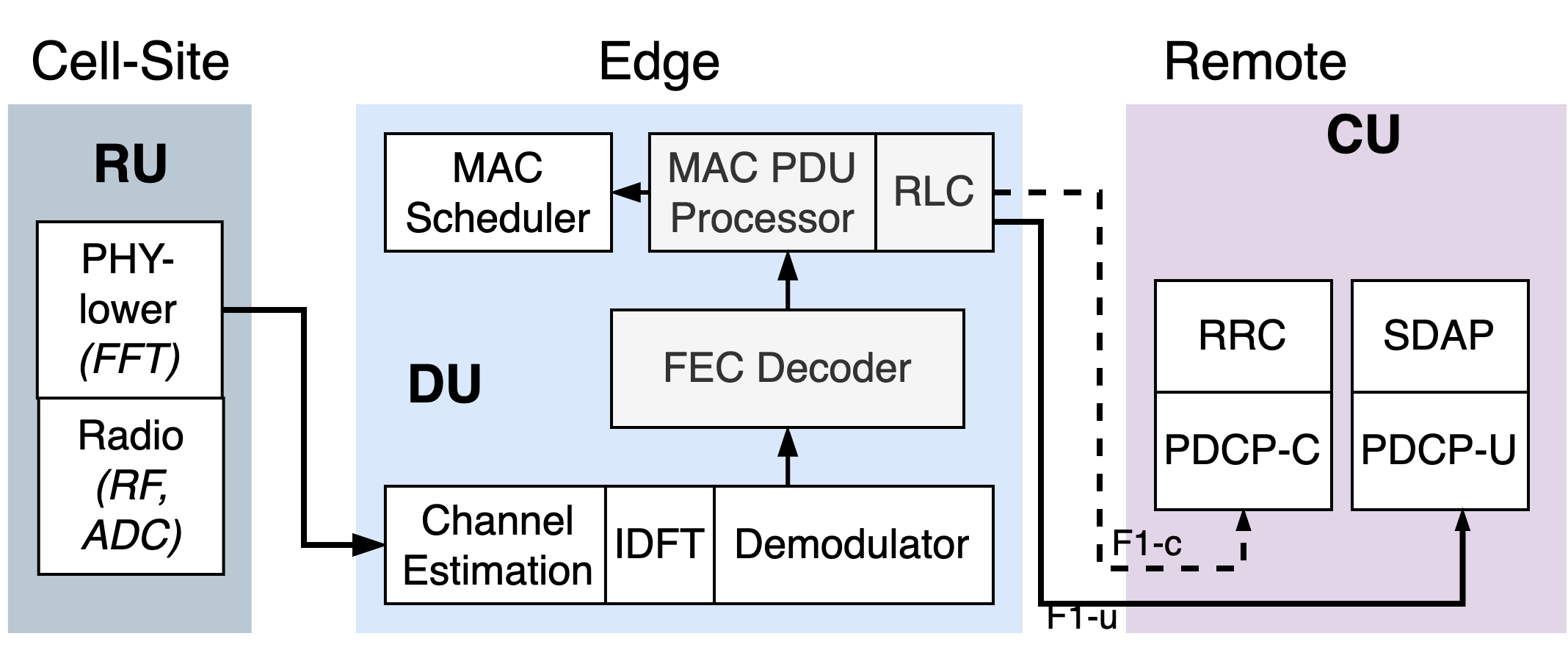}
  \caption{Current vRAN deployments: RU at cell sites, DU at edge computing facilities, and CU at remote/centralized computing facilities.
  }
  \label{fig:topo}
\end{figure}

Traditional radio access network (RAN) realizations relied on proprietary hardware and software (i.e., baseband processing units (BBUs)) for performance and reliability, deployed near base stations (or cell towers) according to the required cell capacity. Relentless technology ``softwarization'' and virtualization trends have resulted in virtualized RAN (vRAN) architectures, enabling ``split'' BBU deployments, with some functionality executing on common-off-the-shelf (COTS) compute hardware using distributed topologies, as shown in Figure~\ref{fig:disttopo}.
vRAN deployment splits the RAN functionality into a radio unit (RU), a distributed unit (DU) and a centralized unit (CU)~\cite{ORAN} .\footnote{Strictly speaking the O-RAN architecture refers to these as O-RU, O-DU and O-CU respectively.}
As shown in Figure~\ref{fig:topo}, RUs are located at cell sites (distributed throughout the deployment area) and connected to DUs at edge compute locations via a fronthaul network. DUs, in turn, are connected to CUs located at remote/centralized compute facilities via a midhaul (MH) network. The F1 interface enables this disaggregation for both control (F1-C) and user plane (F1-U) communication. vRAN aims to realize RAN functionality (for the DU and CU) in software on general-purpose compute platforms (as opposed to proprietary hardware). As with other virtualization efforts, i.e., cloud computing and network function virtualization, the expectation is that vRAN technology will result in a reduction of the total cost of ownership with unified hardware and software infrastructure, simplified network operation and maintenance through unified management, and, fostering flexibility and innovation~\cite{IntelvRAN,RT-OPEX,CloudIQ}.\footnote{Indeed vRAN architectures are seeing real-world deployments from service providers such as Rakuten, Dish and Verizon.}

\begin{table*}
    \centering
    \caption{Example deployment and resources at different locations for distributed vRAN}
    \label{tab:edge_remote}
    \begin{tabular}{ccccc}
        \toprule
        \thead{Location \\(Device)} & 
        \thead{Function} & 
        \thead{Space Cost \\(Relative)} & 
        \thead{Distance/Latency} & 
        \thead{Bandwidth} \\
        \midrule
        Cell-Site (RU) & RF frontend & Very high & -- & -- \\
        Edge (DU) & PHY, MAC, RLC & High & $<10\,\text{km} / <500\,\mu\text{s}$ & Very high \\
        Remote (CU) & PDCP, RRC/SDAP & Low & $<40\,\text{km} / <10\,\text{ms}$ & User bandwidth \\
        \bottomrule
    \end{tabular}
\end{table*}

\begin{figure}
  \centering
  \includegraphics[width=\linewidth]{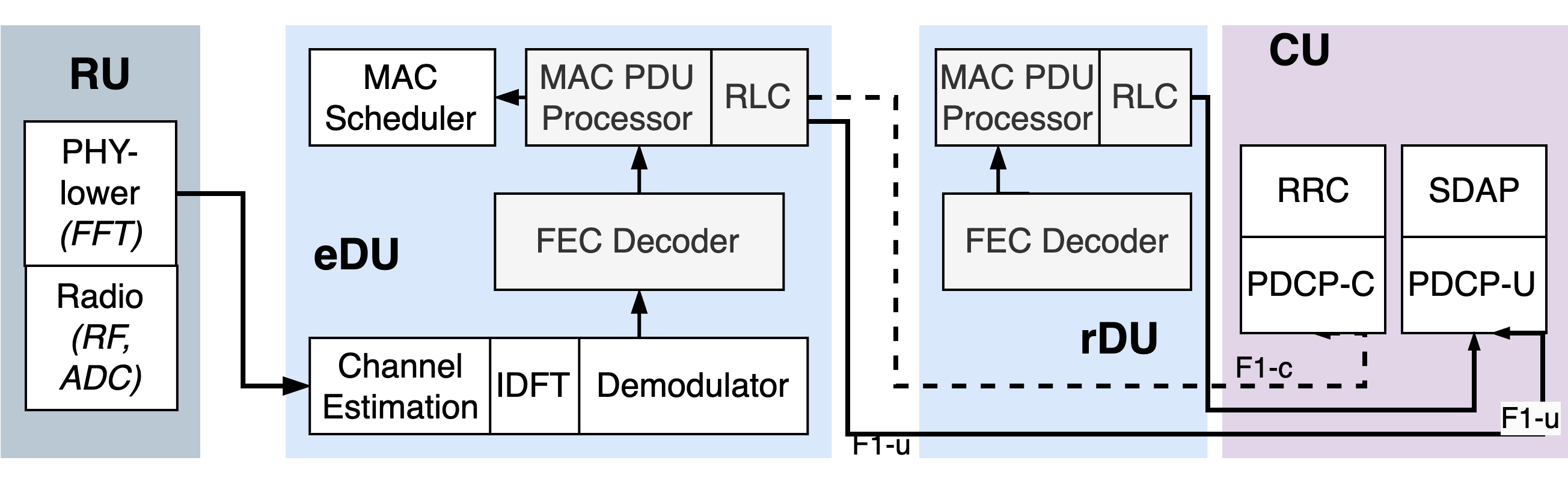}
   \caption{\Name{}'s split-DU (edge DU(eDU) and remote DU(rDU) architecture}
  \label{fig:HadesStack}
\end{figure}

In practice, however, the potential benefits of a virtualized/cloud-native RAN are hindered by 
\textit{the high costs of over-provision of edge computing resources to support the DU}.
Table~\ref{tab:edge_remote} outlines the space, distance/latency, and bandwidth requirements of a vRAN deployment. The strict latency requirements place low-layer RAN functions (PHY, MAC, RLC) at the edge, leading to significant computational overhead, particularly for uplink decoding, which can account for over 60\% of the vRAN workload~\cite{RT-OPEX,CloudIQ,foukas2021concordia,Nuberu}.
Edge capacity is further strained by the bursty nature of RAN traffic, requiring substantial over-provisioning of CPU resources. Even with workload consolidation, many CPU cycles in vRAN environments remain unused due to micro-scale workload fluctuations~\cite{foukas2021concordia}. Limited edge capacity\footnote{Edge facilities have the smallest available footprint, and the number of servers the edge site can host can range from more than 100 in the largest footprint to as low as three servers in its minimal form~\cite{CiscoReimaging}.}, high failure rates of standard nodes\footnote{According to one study general purpose compute nodes have an annual failure rate of 11\% for four-year-old equipment~\cite{x86_fail}.}, and the need for backup servers increase deployment costs and risk service disruptions. For example, data from Rakuten in Japan show only 99.7\% RAN availability without additional edge compute capacity~\cite{Rakuten}. Insufficient edge capacity leads to ungraceful network degradation during overloads, either by not admitting users or degrading service.

Given this context, we pose a crucial question: \textit{Can we strategically offload a segment of the DU's demanding workload from the edge to remote sites, effectively alleviating the hefty deployment costs at the edge intrinsic to current vRAN implementations?} As we explore this question, we make a number of observations.

\noindent\textbf{First}, 
in 5G, the end-to-end latency requirement of traffic depends on the packet content, and most uplink traffic is \textit{not} latency-sensitive.
For instance, video streaming applications can tolerate end-to-end delay up to 150ms~\cite{zoom}, successfully operating on 4G networks with end-to-end latency often exceeding 100$ms$. Thus, it is reasonable to expect that many applications will continue to function well with increased latency, such as an additional 10s of milliseconds over the middle-haul network.
Notably, control traffic over the F1-c interface, such as layer 3 Radio Resource Control (RRC) messages, can tolerate latencies higher than several hundred milliseconds~\cite{3gpp}.

\noindent\textbf{Second}, 
physical layer signal processing tasks, especially uplink decoding, occupy the majority of the vRAN compute time. Offloading latency-tolerant decoding to remote sites can significantly reduce edge DU deployment costs. Early decodability prediction work shows that only a few decoding iterations are required for HARQ feedback~\cite{EHARQ,FastHarqURLLC}.

\noindent\textbf{Third}, 
the MH network cost is more affordable than edge resource deployment in terms of power and space.\footnote{For example, decoding one cell at the edge needs 24W (a 400W Nokia Airframe server supports 10~cells), while offloading only needs 3.5W of power consumed by an optical module.}
This makes offloading latency-tolerant DU tasks from the edge to remote locations an attractive option.

Based on the above observations, 
we propose investing in possible trade-offs between edge computing resources and networking resources. Specifically, as shown in Figure~\ref{fig:HadesStack}, we introduce \Name{}, a system that splits DU functionality into edge and remote parts, allowing dynamic DU workload distribution to enhance efficiency and elasticity over edge infrastructure resources. \Name{} aims to optimize resource use by dividing the decoding process into a \textit{latency-critical early decoding phase} and a \textit{latency-tolerant completion decoding phase}, adaptively distributing the completion decoding processing between edge and remote clouds. 
This approach enhances uplink capacity while utilizing fewer edge computing resources. However, realizing such a design presents three main challenges, which we address as follows:

\begin{description}[wide, labelindent=0pt]
    \item[Identifying Decoding content]

    In realistic cellular scenarios, traffic with different latency requirements is multiplexed in a single layer-1 transmission block. Control messages need consistently low latency, and low-latency applications (e.g., AR/VR) cannot tolerate additional decoding delays. 
    Unfortunately, for 5G uplink traffic, the content of a packet is unknown by the base station before decoding, as the scheduling grant only specifies the data size, with the data content within a transmission left to the user device. 
    This necessitates a method to \textit{identify content} to decide whether a coding block can be offloaded.
    
    \textit{\Name{} addresses this by employing a pre-parsing method for MAC PDU subheaders, checking "bits of interest"(subheaders bits) during early decoding to identify content without completing full decoding.}

    \item[Spliting DU]
    Naively moving all decoding to the remote location would require all remotely decoded traffic to be forwarded back to the edge. This would not only add latency for upper layer processing, but also increase MH bandwidth requirements. To prevent these inefficiencies, \Name{} employs a split DU design whereby DU functionality is duplicated across the edge and remote locations. 
    This design, however, necessitates meticulous synchronization and consistency management across both locations. 

    \textit{As show in Figure~\ref{fig:HadesStack}, \Name{} features a clean design of the split DU with a separate control plane interface at the edge (F1-c) and new data plane interface at both edge and remote (F1-u).
    This design simplifies DU state management at the edge and avoids additional latency from user data traffic moving back and forth between edge and remote locations}.

    \item[Adaptable utilization of edge resources] 
    The bursty nature of uplink traffic at very fine time  granularities (millisecond)~\cite{foukas2021concordia}, makes it non-obvious how to distribute resources between early decoding tasks and completion decoding tasks and when completion decoding should be processed at the edge or offloaded to the remote location. 
    First, an efficient scheduling policy is needed to utilize edge computing power effectively. Second, naively sending all excess completion decoding of a burst to the remote location can increase communication and computation costs while reducing spectral efficiency. This is especially true in cases where the edge location might have enough compute cycles to spare in the upcoming transmission periods (idle period after traffic burst), with a similar (or even smaller) processing latency penalty. 
    Additionally, the midhaul link used for offloading decoding tasks is shared among cells, so its available capacity might fluctuate, constraining the amount of traffic that can be offloaded. 
    
    \textit{\Name introduces a resource allocation framework that dynamically distributes edge compute capacity and offloads excessive completion decoding to the remote. It applies an Earliest Deadline First (EDF) policy at the edge for scheduling, achieving low latency and fair resource sharing. \Name decides which packets to offload based on the edge's spare capacity.}

\end{description}


We use the open-source srsRAN~\cite{srsRAN} as the code base for the implementation of \Name{}.
We evaluate \Name{} with both over-the-air single-cell setup and aggregated multiple-cell-emulated traffic setup. 
The results show that \Name{} can significantly improve the efficiency and elasticity of edge infrastructure resources.
\Name{} has almost no decoding capacity decrease with only half of the edge resource needed to meet the RAN latency for eMBB services. 
\Name{} also proves to be particularly useful/elastic under variable edge and/or network conditions where available edge capacity is scarce.
It maintains 78\% of the maximum decoding throughput even in cases of no available edge decoding resources.

%% file: relatedworks.tex
\section{Related work}
\begin{table*}
  \caption{\coloredtext{red}{Comparison of related work in vRAN }\\
  }
  \label{tab:related}
  \scalebox{1}{
      \begin{tabular}{ccccl}
        \toprule
        \vtop{\hbox{\strut Approaches}} 
        &\thead{Offloading}
        &\thead{HARQ prediction}
        &\thead{PreParsing}
        &\thead{Resource pooling}\\\\        
        \midrule
        \Name{}     &Yes  &Proactive(~\ref{early_decoding})      &High accuracy &Yes\\
        Nuberu    &No   &Passive        &No &Yes\\
        RT-opex   &No   &No             &No &Yes\\
        CloudIQ   &No   &No             &No &Yes\\
        \bottomrule
      \end{tabular}}
\end{table*}

The concept of Virtualized RAN (vRAN) has garnered significant interest in recent years~\cite{CiscoReimaging, IntelvRAN}. The role and impact of virtualization in RAN systems have been thoroughly explored~\cite{ORAN}. A primary challenge in deploying compute-intensive RAN/vRAN processing over cloud-based platforms is meeting the real-time requirements of the PHY layer, especially for computationally demanding tasks like FEC decoding. Various strategies have been developed to manage the allocation of compute resources for these real-time constraints in RAN/vRAN functions.

In the commercial realm, dedicated accelerators are often employed to support RAN functions, notably FEC decoding. Examples include Intel FlexRAN~\cite{flexran} and NVIDIA Aerial~\cite{Aerial}, both of which offer Layer-1 (L1) solutions in commercial vRANs. These systems utilize specialized hardware, such as FPGAs in FlexRAN and GPUs in Aerial, to boost performance. However, implementations details on these platforms are limited as they are not publicly available as open-source.

Academic research has also been active in exploring methods to enhance the efficiency of real-time RAN processing to minimize resource usage. For instance, CloudIQ~\cite{CloudIQ} adopts a scheduling approach that assigns a set of base stations to a computing platform based on their processing demands to fulfill real-time processing needs. RT-OPEX~\cite{RT-OPEX} dynamically redistributes parallel tasks, including decoding, to idle computing resources in real-time, optimizing the use of edge computing. More recently, vrAIn~\cite{vrAIn} introduced machine-learning-based controllers to manage the distribution of radio and compute resources across sliced RAN instances, although it struggles to maintain RAN function reliability during computing resource shortages.

These works typically treat the entire decoding process as constrained by real-time requirements, overlooking the potential benefits of additional decoding time afforded by decoding result predictions (HARQ predictions). Nuberu~\cite{Nuberu}, a closely related work focused on LTE, passively uses HARQ prediction to prevent deadline misses and limits spectrum assignments to reduce load during edge resource scarcity. However, it still relies entirely on edge resources for decoding, thus restricting its capacity.
\coloredtext{red}{
In contrast, as shown in Table~\ref{tab:related}, \Name{} introduces an innovative decoding scheduling framework that actively uses HARQ prediction and pre-parsing for packet subheaders. It intelligently manages resource allocation between edge and remote servers, enhancing overall system efficiency and capability.
}

%% file: background.tex
\section{Background}


\label{sec_background_primes}

In 5G networks, radio activities are organized into defined Transmission Time Intervals (TTIs) or slots, such as $1ms$ for eMBB service. The orchestration of RAN resources involves a complex blend of allocation, transmission, feedback, and control.

\begin{figure}
     \centering
     \begin{subfigure}[b]{0.48\textwidth}
         \centering
         \includegraphics[width=\linewidth]{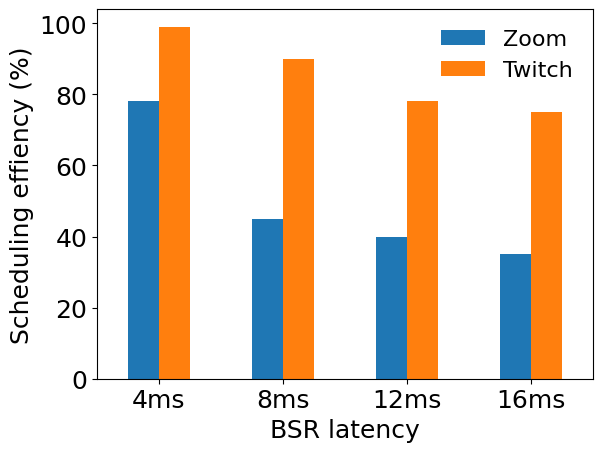}
         \caption{Scheduling efficiency (BSR latency)}
         \label{fig:BSR_latency}
     \end{subfigure}
     \hfill
     \begin{subfigure}[b]{0.48\textwidth}
         \centering
         \includegraphics[width=\linewidth]{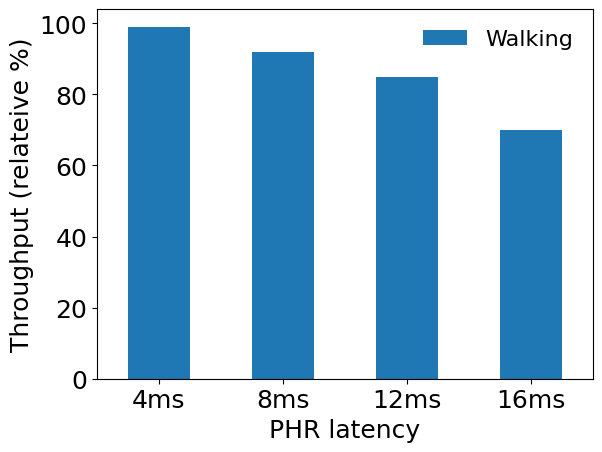}
         \caption{Throughput decrease (PHR latency)}
         \label{fig:PHR_latency}
     \end{subfigure}
        \caption{RAN performance impacted by increased latency in MAC-CE processing} 
        \label{fig:BSRPHR_latency}
\end{figure}

\begin{figure}
  \centering
  \includegraphics[width=0.85\linewidth]{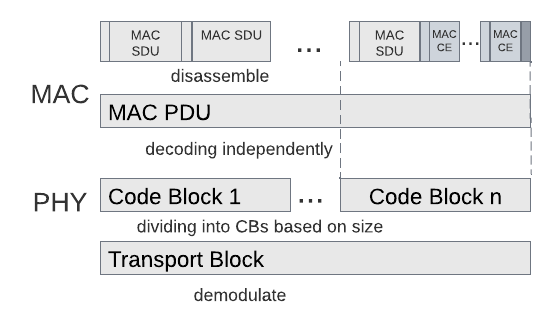}
  \caption{MAC-CE and MAC SDU are wrapped at the end of a MAC PDU, has to be decoded at the edge to guaranteeing low latency of MAC control messages.}
  \label{fig:MACCE}
\end{figure}

\begin{description}[wide, labelindent=0pt]
\item[Uplink Scheduling Efficiency:] 
A critical aspect of RAN efficiency is the proficient scheduling of Physical Resource Blocks (PRBs) and Modulation and Coding Schemes (MCS). PRBs, composed of several subcarriers within a TTI, are the basic resource units assigned to users. MCS, which determines the data rate per PRB, is selected based on user feedback like Buffer Status Report (BSR), Power Headroom Report (PHR), Quality of Service (QoS) requirements, and Channel State Information (CSI) assessments. After PRB allocation and MCS determination, the base station provides a scheduling grant for uplink data transmission.
A key mechanism in this process is the Hybrid Automatic Repeat Request (HARQ) protocol. If decoding issues arise, indicated by a NACK, retransmission is prompted. Delays in HARQ feedback processing can lead to unnecessary retransmissions, wasting resources~\cite{Nuberu, srsRAN, OAI}. Insufficient compute resources, cause high Block Error Rates (BLER), which impact MCS selection. The 3GPP standard aims for a BLER below 5\%~\cite{3gpp}.

\item[Importance of MAC-CE Processing:]
MAC Control Elements (MAC-CEs), including BSR and PHR, are crucial for MAC scheduling. Increased latency in processing these elements can significantly degrade network performance. This impact was demonstrated in an srsRAN-based experiment, with results shown in Figure~\ref{fig:BSRPHR_latency}. By intentionally delaying MAC-CE processing, we observed its effect on RAN performance. Delays in BSR processing led to inefficient resource allocation and higher packet latency. Similarly, delayed PHR processing, particularly for mobile users, directly reduced cell throughput. This was evident as mobile phones at the edge of a base station's coverage experienced decreased throughput due to delayed PHR processing.
\item[MAC PDU Decoding:] 
Uplink transmission entails the mobile device assembling a MAC PDU packet based on the resource assignment from the base station. Logical Channel IDs (LCIDs) within sub-PDU headers identify the SDUs contained, guiding the base station in extracting relevant data for further action or processing. This packet is then decoded by the base station, extracting MAC SDUs for higher-layer processing and MAC-CEs for control functions. If the packet size exceeds 8448 bits, it's segmented into multiple code blocks for independent decoding~\cite{3gpp}, as shown in Figure~\ref{fig:MACCE}. Both 5G NR and LTE employ iterative decoding algorithms~\cite{turbo, LDPC}, traditionally completing all iterations before generating HARQ feedback. Due to the unpredictable and computationally intensive nature of this process, over-provisioning of computational resources is a common practice to ensure low latency and reliable decoding. Decoding error prediction, which involves generating HARQ feedback based on early prediction rather than after complete iterations, is used to enhance performance and manage computational resource constraints~\cite{Nuberu,FastHarqURLLC}. This approach, is especially crucial in using Log-Likelihood Ratio (LLR) based methods, optimizing by assessing the likelihood of each transmitted bit (being '0' or '1') based on extrinsic information during decoding~\cite{EHARQ}, thus predicting decoding results early in the process.

\end{description}

%% file: design.tex
\section{\Name{} Design}
The essence of \Name{} lies in its innovative approach to uplink decoding, which is divided into two separated phases: "Early decoding" and "Completion decoding." Early decoding operates within a narrow time budget (e.g., 1ms) and is executed at the edge due to its time-sensitive nature. Completion decoding, on the other hand, has a more flexible time budget and can either be processed at the edge or offloaded to remote servers based on its latency requirements. This separation is facilitated by a split Distributed Unit (DU) architecture, which enables effective vRAN operations. \Name{} further enhances operational efficiency through a sophisticated framework for scheduling decoding tasks and managing offloading processes.

In this two phase decoding strategy, we use task queues for managing the decoding flow. Specifically, incoming packets are initially forwarded to the early decoding queue, where they undergo decodability prediction and pre-parsing of the MAC PDU. The subsequent forwarding of these packets to specific completion decoding queues is based on their identified content and respective decoding delay budgets. This approach allows for packets with longer delay tolerances to be potentially offloaded to remote servers if the edge capacity is exceeded.

\subsection{Decoding splitting}

\begin{figure*}
     \centering
  \includegraphics[width=\linewidth]{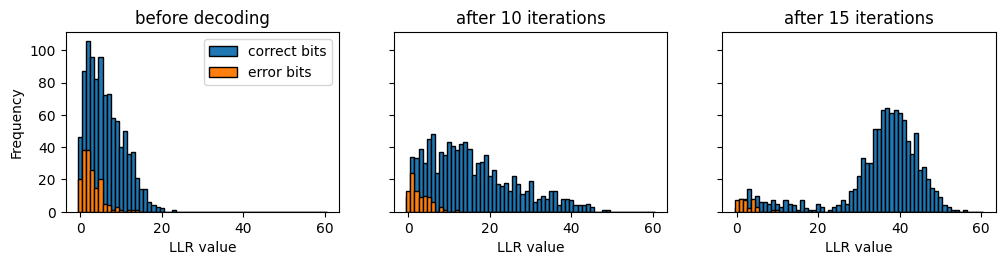}
  \caption{LLR distribution of error/correct bits of an example decoding block}
  \label{fig:LLR_distribution}
\end{figure*}

\subsubsection{Early decoding}\label{early_decoding}
The early decoding phase in \Name{} serves two primary functions: (1) predicting the decodability of a Transport Block (TB), and (2) initiating the pre-parsing of subheaders within the MAC PDU encapsulated in that TB.

The predictive capability of early decoding stems from an analysis of the average Log-Likelihood Ratio (LLR) values across early decoding iterations, which can indicate the likelihood of successful decoding before the final CRC check~\cite{EHARQ,LLRhspa,FastHarqURLLC}. This approach relies on the observation that extrinsic information, represented by LLRs, converges through successive iterations if the packet is decodable. Such convergence allows the base station to potentially pause the decoding early before CRC passing or the reaching of the maximum number of iterations. Conversely, a lack of convergence in the LLRs would suggest an undecodable packet.

In \Name{}, each packet's early decoding phase integrates with decodability predictions, as shown in Algorithm~\ref{alg:Prediction}, to determine if further iterations are required. If the outcome remains uncertain (low LLRs), the process iterates until either the maximum iteration limit or the HARQ deadline is met. 
We have two threshold vectors ($\vec{A}=[a_{0} \dotsm a_{m}]$, $\vec{N}=[n_{0} \dotsm n_{m}]$) that are used to predict the decodablity.
This proactive prediction strategy contrasts with a prior passive method~\cite{Nuberu} that defers prediction until a decoding deadline is imminent, thus missing the opportunity for flexible scheduling based on deadline constraints

Predicting decodability by itself is insufficient in a realistic RAN setting with diverse QoS requirements and MAC control loops. Treating all packets and bits equally fails to address the unique demands of latency-sensitive traffic and can disrupt MAC control loops. To address this, \Name{} enhances early decoding by not only predicting decodability through average LLR-based estimation but also by pre-parsing MAC subheaders to identify decoding content and MAC Control Elements (CEs) (to this end, we focus on BSR and PHR).

We assess the correlation between the evolution of individual LLR values over iterations and the accuracy of their corresponding bits. Figure~\ref{fig:LLR_distribution} shows our analysis of a 1000-bit code block across 20 iterations and demonstrates the ability to utilize specific LLR values as indicators for bit correctness.
For packets that fails in CRC checking before final completion of decoding, the correctness of a single bits can still be estimated based on its single LLR value.
By examining the subheaders in an incompletely decoded packet, it is feasible to identify the PDU contents by extracting LCGIDs bits. Although the packet may not be fully decoded without passing CRC checking, the bit error rate post-prediction iterations is markedly lower. Given that headers represent a small fraction of the entire packet, we can achieve high accuracy in detecting subheaders for each sub-PDU.

Unlike the single-header structure of LTE MAC PDUs, the 5G NR MAC PDU contains multiple subheaders, each preceding a MAC subPDU and providing the information necessary for decoding that subPDU.
We devise an early parsing algorithm, Algorithm~\ref{alg:EarlyParsing}, specific to the 5G architecture, where each header includes a length field that indicates the position of subsequent subPDUs. Notably, MAC-CE subPDUs are positioned at the end of the MAC PDU. The algorithm scans from the beginning to the end of the PDU, retrieving headers to locate the MAC-CE headers via the "LCID" field. Packets with discerned MAC-CEs are expedited for decoding to ensure minimal latency. A threshold $L_t$ is used to estimate the correctness of each bit. 

\begin{algorithm}
\scriptsize 
\SetAlgoLined
\KwResult{Decodability Prediction for LDPC Codeblock}
\textbf{Input:} LLR values of the codeblock, Maximum iterations $I_{MAX}$, Threshold vectors $\vec{A}$, $\vec{N}$ for prediction\;
\textbf{Output:} Decodability Status\;
\textbf{Initialize:} $i \gets 0$, Decodability Status $\gets$ \textit{UNKNOWN}\;

\While{$i < I_{MAX}$ \textbf{and} Decodability Status $=$ \textit{UNKNOWN}}{
  $i \gets i + 1$\;
  Get $L_i$ with one more decoding iteration\;
  \For{each LLR value $L_i$ in the codeblock}{
   \uIf{$L_i > a_i$}{
    Decodability Status $\gets$ \textit{DECODABLE}\;
    \textbf{break}\;
   }
   \uElseIf{$L_i < n_i$}{
    Decodability Status $\gets$ \textit{NOT DECODABLE}\;
    \textbf{break}\;
   }
  }
  \uIf{Decodability Status $=$ \textit{UNKNOWN} \textbf{and} $i = I_{MAX}$}{
   Decodability Status $\gets$ \textit{NOT DECODABLE}\;
  }
}
$i_{prediction} \gets i$ \;
\caption{Algorithm of decodability prediction}\label{alg:Prediction}
\end{algorithm}

\begin{algorithm}
\scriptsize 
\SetAlgoLined
\SetKwInput{KwInit}{Initialize}
\KwInit{$i \gets i_{prediction}$ \tcp*{initial iteration index is set for prediction}}
\SetKwFunction{FParsingSubHeader}{ParsingSubHeader}
\SetKwProg{Fn}{Function}{}{end}
\Fn{\FParsingSubHeader{$\vec{LLR}$, $Position_s$}}{
   \uIf{all $l_k > L_t$ for each bit $k$ in $Position_s$}{
      Extract $LCID_s$, $L_s$ \; \tcp{Get LCID and length of sub-PDU $s$}
      \Return $True$\;
   }
   \Else{
      \Return $False$\;
   }
}

\While{$i < I_{MAX}$ \textbf{and} CRC check not passed}{
   $i \gets i + 1$; $s \gets 0$; $Position_s \gets 0$\;
   Get $\vec{LLR_i}$ with one more decoding iteration\;
   \While{not reached the end of PDU}{
      \uIf{\FParsingSubHeader{$LLR_i$, $Position_s$}}{
         $s \gets s + 1$\;
         $Position_s \gets Position_{s-1} + L_s$\;
      }
      \Else{
         \textbf{break} \;
         \tcp{Terminate the loop in the case when subheader parsing fails}
      }
   }
}
\caption{Algorithm for pre-parsing in early decoding}\label{alg:EarlyParsing}
\end{algorithm}

\subsubsection{Completion decoding}
After early decoding extracts critical bits and determines decoding status, additional iterations are needed for completion. The goal of this phase is to finish decoding packets and ensure every bit is correctly decoded (passing PDU CRC checking). To enable QoS differentiation for traffic streams and computation efficiency, we have separate queues for different delay budgets. PDUs are pushed into the queues based on the content with the lowest delay budget within the PDU. The schedculing and offloading details are provided in~\ref{scheduling framework}.

\subsection{Split DU Architecture in \Name{}}
\coloredtext{red}{
To support decoding splitting, \Name{} incorporates a comprehensive DU split design, as depicted in Figure~\ref{fig:splitDU}. This design includes the splitting and/or duplication of DU stack components between the edge and remote sites, as well as the management of DU states and interfaces. The design leverages the inherent strengths of edge and remote processing, ensuring both efficiency and low latency.
}

\subsubsection{PHY component}
\coloredtext{red}{
Low PHY components such as channel estimation, FFT, and demodulation are localized at the edge. This ensures that upstream processing remains uninterrupted, leveraging the latency advantages of local processing. By keeping these critical components at the edge, \Name{} can guarantee quick and efficient initial processing of incoming data.
In term of decoding, the early decoding phase is executed at the edge to quickly determine the decodability of Transport Blocks (TBs) and initiate pre-parsing of subheaders within the MAC PDU. If early decoding suggests that a TB is likely to be decodable, further processing can be deferred to the completion decoding phase, which may occur at the remote site or edge site depending on resource availability and latency requirements. For example, for AR/VR traffic with required low latency, completion decoding also remain at the edge. While for video traffic that can tolerate additional MH latency, it's completion decoding can be offloaded if needed.
}

\subsubsection{MAC component}
\coloredtext{red}{
Building on the split decoding, the split design extends to the MAC layer.
The MAC scheduler is positioned at the edge due to its latency-sensitive nature. It takes decodability prediction results and MAC CE bits from pre-parsing as input for scheduling. Placing the MAC scheduler at the edge ensures that scheduling decisions can be made swiftly, enhancing the overall responsiveness of the network.
The states that the MAC scheduler relies on are maintained solely at the edge. This centralized state management reduces synchronization overhead and simplifies the overall system design.
}

\coloredtext{red}{
In the early decoding phase, MAC-CEs are identified through pre-parsing and extracted with the lowest delay budget. This necessitates distinguishing the MAC-CE sub-PDU processor from the general MAC PDU processor. By processing MAC-CEs promptly, \Name{} ensures that control loops remain efficient and responsive.
The MAC PDU processor, a stateless function that does not depend on user or cell context stored in DU states, is replicated at both the edge and remote sites. This setup enables immediate processing of PDUs and efficient multiplexing of traffic streams after completion decoding. The replication of the MAC PDU processor facilitates a seamless transition between edge and remote processing, ensuring continuous and efficient handling of data packets.
}

\subsubsection{RLC Components}
\coloredtext{red}{
These entities manage functions such as segmentation, reassembly, and error correction to ensure reliable data transmission.
RLC entities are deployed at edge or remote sites for different traffic flows. It would be straightforward to deploy RLC entities based on the traffic latency and its upstreaming decoding process.
For critical traffic flows, RLC entities are deployed at the edge to minimize transmission delays and ensure timely data delivery. And for traffic flows that are less sensitive to latency, RLC entities are deployed at the remote site.
}

\subsubsection{Message Flows}
\coloredtext{red}{
Based on the placement of components of split DU, message flows are orchestrated accordingly. 
RRC messages are routed back to the edge, maintaining a cohesive F1-C interface with the Centralized Unit (CU). Concurrently, SDUs of user traffic with high latency budget are directed to the remote site, interfacing with the CU via a separate F1-U channel that consolidates user traffic. 
}

\coloredtext{red}{
This split DU arrangement ensures a clear demarcation of F1 interfaces between the split DUs and the CU, facilitating harmonious and efficient system operation. By streamlining message flows and interface management, \Name{} achieves a adaptable and optimized network architecture.
}

\begin{figure}[h!]
    \centering
    \begin{subfigure}[t]{0.9\textwidth} 
        \centering
        \includegraphics[width=\linewidth]{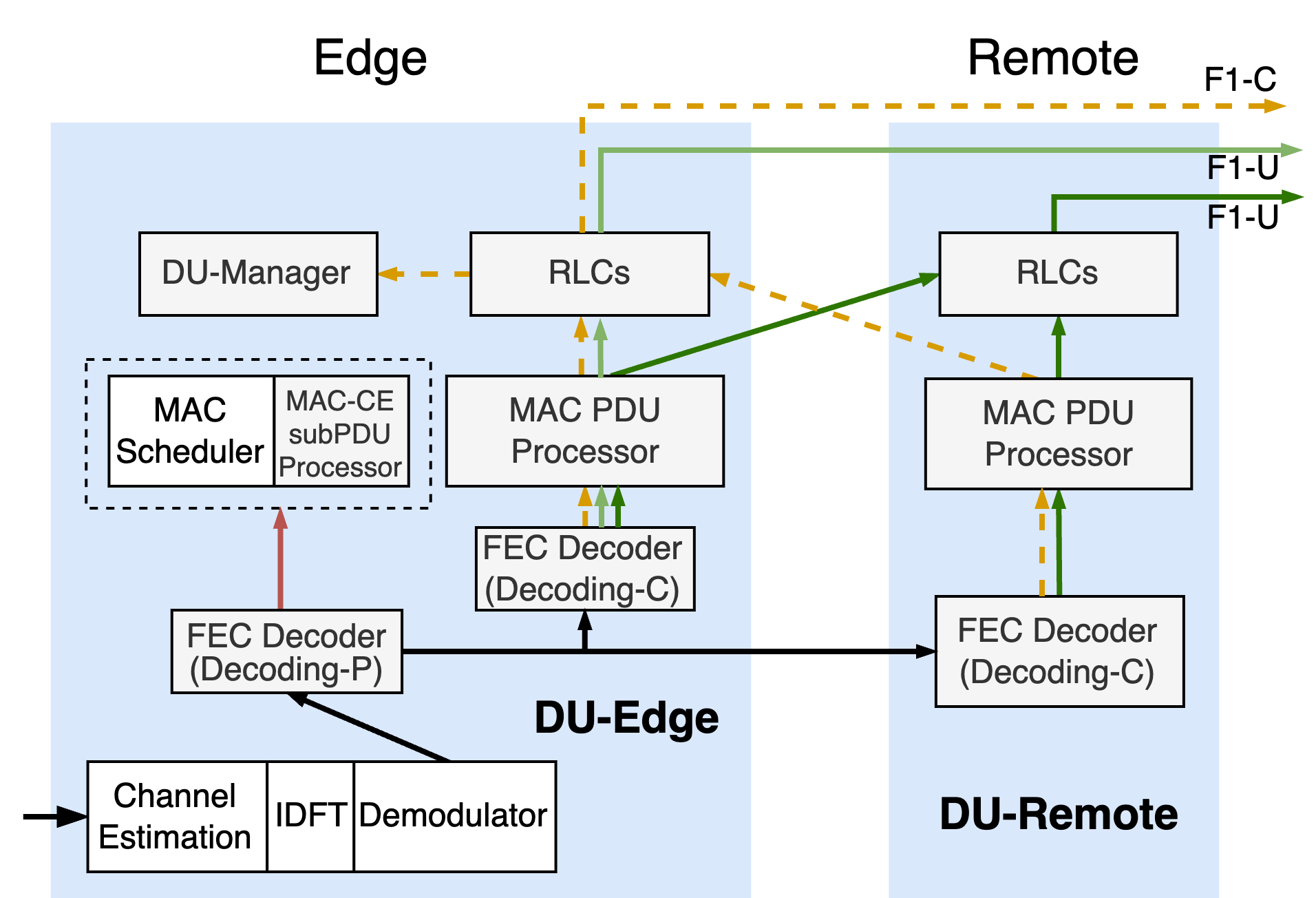}
        \caption{\coloredtext{red}{Split DU architecture of \Name{}}}
        \label{fig:splitDU}
    \end{subfigure}
    \hfill
    \begin{subfigure}[t]{0.9\textwidth} 
        \centering
        \includegraphics[width=\linewidth]{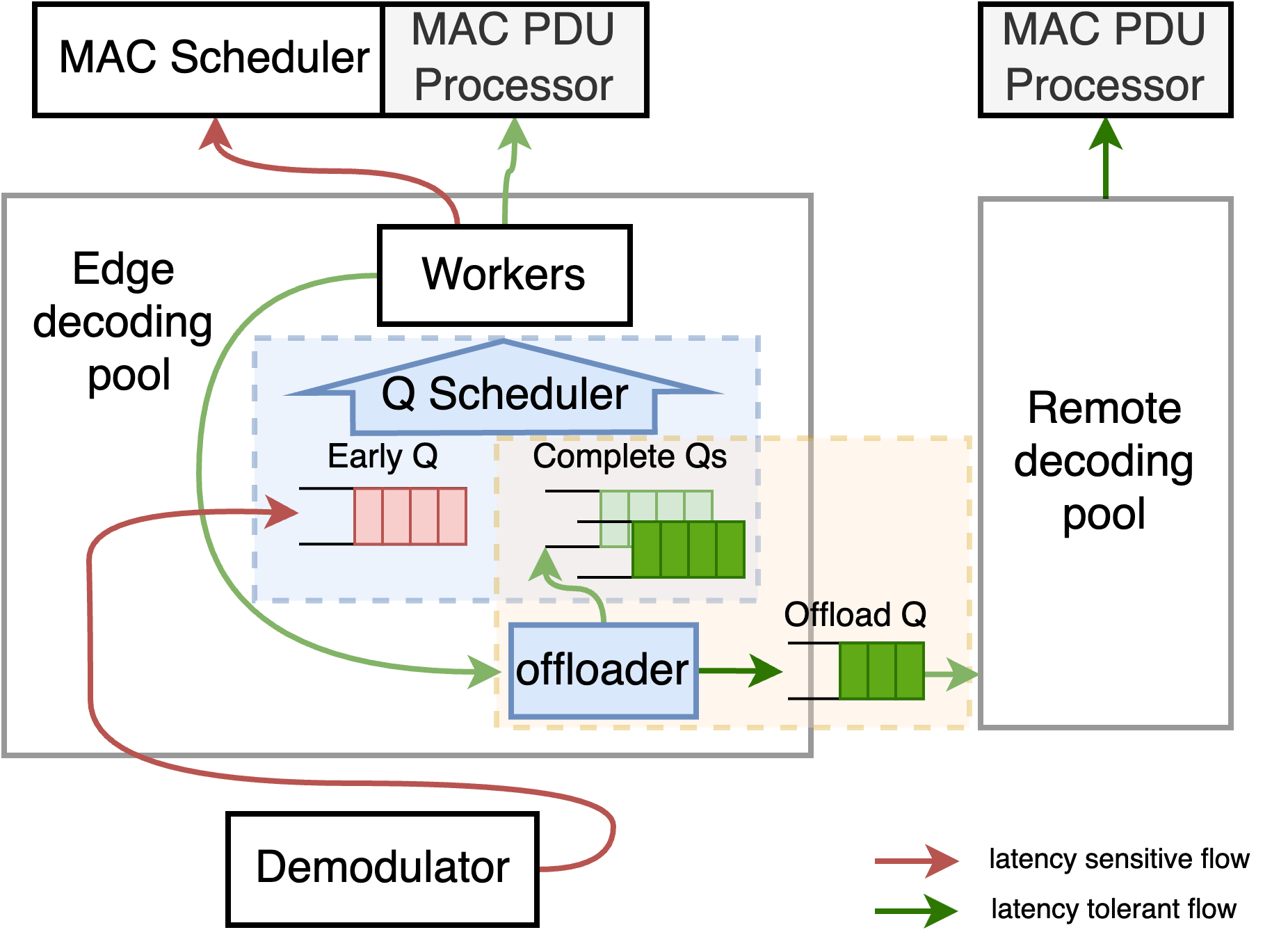}
        \caption{\coloredtext{red}{\Name{} framework of scheduling and offloading}}
        \label{fig:scheduling}
    \end{subfigure}
    \caption{Overall architecture and framework of \Name{}} 
    \label{fig:architecture}
\end{figure}

\label{scheduling framework}
\subsection{Decoding scheduling framework}
The \Name{} design includes a scheduling framework to maximize the efficiency of edge computing and optimize the use of the Middle-Haul (MH) bandwidth. Illustrated in Figure~\ref{fig:scheduling}, this framework includes two interacting schemes: orchestrating decoding tasks at the edge and managing the offloading of completion decoding to remote servers. The design of this scheduling and offloading mechanism are based on several requirements.

First, the framework should prioritize early decoding tasks due to their strict latency constraints, essential for meeting HARQ deadlines and timely MAC control actions. Missing these early decoding deadlines can lead to packet delivery failures and increased Block Error Rates (BLER), which in turn reduce throughput as BLER impacts link adaptation mechanisms. Delays in processing MAC control messages can trigger a cascade of negative effects on RAN performance, including inefficient spectral resource usage, scheduling latency breaches, and inaccurate power allocations.

Second, the framework should account for the latency budget allocated for completion decoding, which is dependent on the specific content within the PDU packet. Inefficient handling of this phase, especially failing to complete decoding within the allocated time, results in squandering computational resources expended in the early decoding stages.

Third, it should carry a discerning offloading policy that identifies the most opportune moments for migrating completion decoding tasks to remote servers. In other worrds, it should ensure an efficent usage of edge computing, aligning the immediate requirements of latency-sensitive decoding with the effective utilization of MH bandwidth, all aimed at minimizing overall latency.

In this context, \Name{} introduces an edge queue scheduling strategy that dynamically distributes computational resources between early and completion decoding phases and an offloading policy that can further dynamically migrate completion decoding load to the remote.

\subsubsection{Scheduling at edge}
The goal of decoding task scheduling at the edge in \Name{} is to balance the load between early and completion decoding, while interacting with BLER-based link adaptation to adjust link load. I.e., the key idea is to use BLER-based link adaptation for load control.
When there are sufficient CPU cores at edge, early decoding tasks should be prioritized to avoid being blocked by completion tasks, preventing missed prediction deadlines and unnecessary load reduction by link adaptation. Conversely, when computing resources are scarce, allowing more missed deadlines with NACKs (uncertain prediction outputs) can lift the BLER above the target threshold (5\%), triggering BLER-based load/link adaptation to reduce the decoding load.

\Name{} employs the Earliest Deadline First (EDF) scheduling policy to manage decoding tasks at the edge. Multiple queues buffer decoding tasks based on their latency budgets. The early queue, with a small decoding time budget (e.g., 1 ms), handles low-latency early decoding tasks, while multiple completion queues handle tasks with varying latency budgets (e.g., 5-100 ms for different QoS levels). Within each queue, tasks are ordered by their remaining time budget, with the task having the shortest remaining time budget at the head of the line.
Tasks are scheduled based on the remaining time budget before their deadlines. The scheduler compares the tasks at the heads of the different queues and schedules the one with the earliest deadline. When there are enough idle CPU cycles, completion decoding tasks are dequeued quickly, ensuring that early decoding tasks are prioritized when the head-of-line task has more than 1 ms remaining. Conversely, when CPU cores are scarce, completion tasks may have less than 1 ms remaining and are thus prioritized.

This dynamic adjustment ensures optimal prioritization and efficient use of computational resources, balancing the need to meet early decoding deadlines with the overall decoding load. By integrating link adaptation, EDF scheduling dynamically adjusts to the edge's decoding capacity, ensuring effective management of decoding tasks.
In other words, the pressure of completion decoding load will increase the early decoding missing rate, which can trigger link adaptation to reduce the decoding load.
\Name{} is able to maintain the BLER of 5\% with dynamic traffic and maximize the utlization of edge decoding capacity.

\subsubsection{Offloading to remote}
\label{offload}
\coloredtext{red}{
To alleviate the pressure on completion decoding loads, \Name{} implements an adaptive offloading control algorithm that dynamically balances the distribution of completion decoding tasks between edge and remote locations. This algorithm employs dynamic offload queuing control to maintain a balance between Middle-Haul (MH) bandwidth usage and decoding latency. 
Overly aggressive offloading can lead to increased demands on MH bandwidth, while insufficient offloading can cause congestion at the edge, resulting in higher latency due to backlogs in the completion decoding queue.
}

\coloredtext{red}{
In \Name{}, a single offloading queue buffers packets destined for remote decoding. This mechanism helps efficiently manage the flow of data between the edge and remote servers. The offloading criteria are based on the queuing delay within the offloading queue and the completion decoding queues. The following equation encapsulates this approach:
}

\begin{equation}
\label{eq_off}
Offload_{i}^{q} = 
\begin{cases}
1, & \text{if } L_{i}^{o}-L_{i}^{MH} < L_{i}^{q} \\
0, & \text{if } L_{i}^{o}-L_{i}^{MH} \geq L_{i}^{q}
\end{cases}
\end{equation}

\coloredtext{red}{
At each defined period $i$, the criteria are computed where $L_{i}^{o}$ denotes the queuing latency of the offloading queue and $L_{i}^{q}$ represents the queuing latency of the completion queue $q$.
\Name{} measures the Head of Line queuing latency of all the queues based on timestamps created when early decoding tasks arrive at the edge DU. The equation signifies that when the difference between the queuing latency for offloading and the queuing latency for edge completion decoding is less than the MH latency $L_{i}^{MH}$, offloading is initiated (indicated by 1) for queue $q$. Conversely, if this condition is not met, offloading is not performed (indicated by 0). This strategic decision-making ensures optimal resource utilization and maintains system efficiency.
To estimate MH latency, we measures the round-trip time (RTT) of the offloaded decoding packet along with its decoding feedback from remote to edge.
}

%% file: evaluation.tex
\section{Implementation}
\Name{} is developed using the srsRAN library~\cite{srsRAN}, leveraging its finely tuned LDPC decoders for X86/ARM server architectures. Validated on x86 platforms, \Name{} can be easily ported to other architectures, such as ARM-based 5G accelerators, with minimal effort. Operating within a Linux environment, \Name{} conforms to the srsRAN’s recommended settings. While designed for 5G NR, its principles allow compatibility with 4G LTE systems, particularly in the context of decoding processes.
The detail of this implementation are explored in the following discussion.

\subsection{Decoding Scheduling}
In available RAN implementations~\cite{OAI,srsRAN}, a slot-based pipeline model for the PHY layer is commonly adopted, offering limited parallelism for vRAN. Specifically, PHY layer processing of sequential slots is pipelined with multiple slot worker threads since the slot arrival period is shorter than the processing time of a single slot~\cite{Nuberu}. To enable the decoding scheduling framework of \Name{}, we separated the decoding worker threads from the slot workers. Each decoder worker is assigned to a single CPU core, allowing the parallelization of decoding tasks across multiple CPU cores. Allocating more cores to the decoder pool increases \Name{}’ decoding capacity (bits/second).

Slot workers pass FEC decoding requests to edge decoder workers through the Early Queue and invoke callback functions when early decoding finishes. Completion decoding is then placed in the completion queues unless it is offloaded to the remote. Each edge worker has its own decoding task queues. For an edge decoding pool of 
$N$ CPU cores, decoding packets are distributed across the 
$N$  cores as fresh decoding tasks arrive. Within a core, the decoder worker executes tasks based on the scheduling of the early decoding queue and completion queues.

\subsection{Offloading Control}
\Name{} employs an adaptation offloading algorithm to dynamically distribute completion decoding tasks between edge and remote locations. Offloading transportation is managed by a dedicated CPU core, with an offload queue shared by edge decoding workers. LLRs and decoding context are transported between edge and remote using a data structure with typed indexes at the headers. Offloading data is managed by a separate thread, utilizing semaphores for thread synchronization.

\subsection{Link Adaptation}
For link adaptation, a timer enforces HARQ deadlines and decode time budgeting. Before each iteration, \Name{} checks the available slack time (the time before a task’s deadline) and determines if the execution time for another decoding iteration is less than the slack time. If the execution time exceeds the slack time, it is identified as a missed deadline. Based on the monitored BLER, the MAC scheduler can adjust the load through link adaptation. The current link adaptation implementation of srsRAN relies on a fixed mapping between expected SNR and MCS for simplicity. We have added a dynamic link adaptation algorithm to the MAC scheduler for \Name{}.

\section{Evaluation}
We evaluate \Name{} using off-the-shelf Intel Xeon Gold 6126 CPU (24 cores @ 2.6 GHz) servers on the edge and remote sides as vRAN servers, connected via a 100 GbE switch. To simulate edge computing resource constraints, we limit the number of cores for edge decoding workers, with each core capable of decoding at 250 bits/µs. Middlehaul (MH) network latency is emulated by adding a virtual queue at the remote server using a virtual interface and qdisc, \coloredtext{red}{
with a typical MH latency setup of 10ms~\cite{mhlatency}, adjustable to measure \Name{}’s performance under varying MH latencies}.

\coloredtext{red}{
We emulated variable decoding loads using multiple 20MHz cells. To test with realistic cell traffic pattens (packet size, MCS, intervals), we used a traffic generator based on traces collected from real-world cells via the Falcon sniffer~\cite{falcon}, capturing a variety of traffic patterns at peak times. This setup records packet size, MCS used, intervals between packets, and traffic follow can be identified with device ID in the grant.
To generate the load, we randomly pick traffic flows from the trace for each cell. The spectrum is fully loaded by adding as many flows as possible.
MCS adaptation for \Name{} is applied to for change packet size with link adaptation. For injecting emulated traffics, emulated radio front-ends are used.
}

It is important to note that the original srsRAN implementation lacks a deadline mechanism for uplink processes. Consequently, any uplink processing delays adversely affect downlink performance, leading to synchronization issues and reduced throughput~\cite{Nuberu}. Due to this inherent limitations, comparing against native srsRAN as a baseline is not possible. Instead, our baseline aligns with common practices observed in commercial RAN deployments for a more fair comparison. Specifically, we consider two non-\Name{} implementations of edge decoders for comparison purposes:

\textbf{BaseLine} has only a decoder pool at the edge with only traditional "run-to-completion" decoders. 
Upon missing the HARQ deadline, unfinished decoding packets are discarded as decoding failures, resulting in NACK feedback for HARQ.

\textbf{Nuberu}-like 
has an edge decoding pool with HARQ prediction and PRB-based load adaptation (or computing congestion controll~\cite{Nuberu})). It cannot offload any decoding loads. Decoding processing at the edge is performed in a traditional run-to-completion fashion, but prediction is allowed when edge decoding resources are deficient. Additionally, traffic is limited by capping the spectrum resource (PRBs) usage for load adaptation.

\subsection{Performance with decoding offloading}
\Name{} is designed to significantly enhance edge efficiency and provide greater elasticity for uplink decoding by leveraging available MiddleHaul (MH) bandwidth. Our evaluation of this part focuses on throughput and latency in a fully loaded UDP traffic scenario over an emulated 10-cell network, each cell operating at 20MHz. We vary the number of CPU cores and available MH bandwidth to assess performance. This controlled environment excludes the influence of MAC Control Elements (MAC-CE) and considers a single traffic type for a fair comparison against the native Nuberu framework, which also does not account for MAC CE and traffic Quality of Service (QoS).

\begin{figure}
\centering
\begin{subfigure}{\textwidth}
  \centering
  \includegraphics[width=\linewidth]{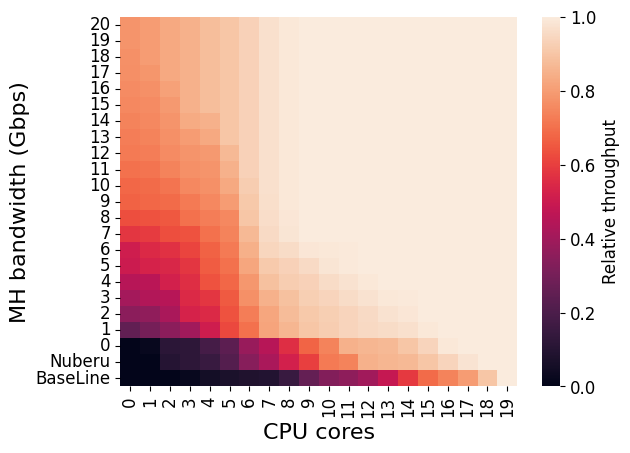}
  \caption{Throughput}
  \label{fig:throughput_heatmap}
\end{subfigure}%
\hfill
\begin{subfigure}{\textwidth}
  \centering
  \includegraphics[width=\linewidth]{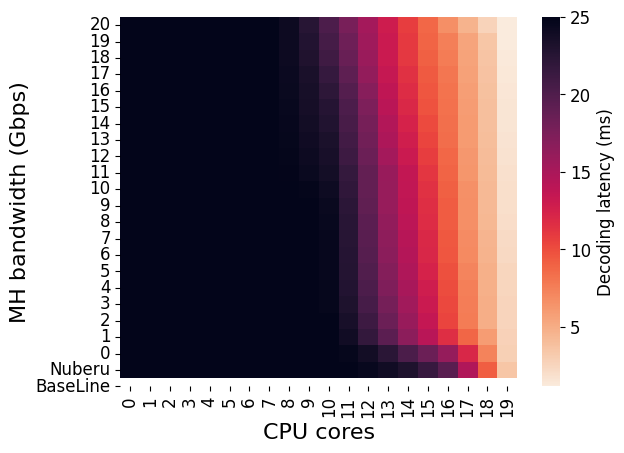}
  \caption{Latency}
  \label{fig:latency_heatma}
\end{subfigure}
\begin{subfigure}{.9\textwidth}
     \centering
  \includegraphics[width=\linewidth]{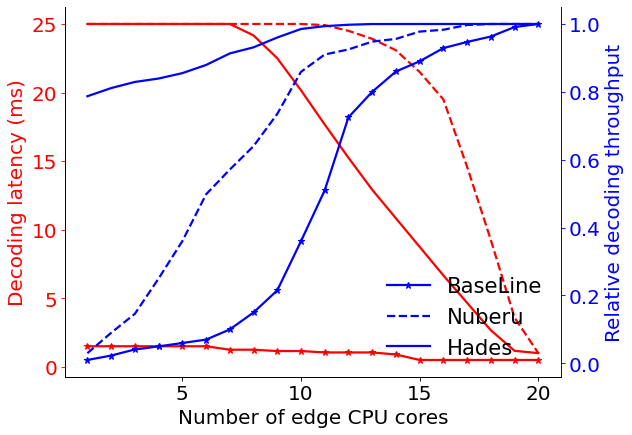}
  \caption{Throughput vs. latency with abundant MH BW}
  \label{fig:throughput-latency}
\end{subfigure}
\caption{\coloredtext{red}{
Performance with limited resource: 10-cells, highlighting \Name{}' superior performance in resource-limited environments with a 20ms packet delay budget}}
\label{fig:heatmap}
\end{figure}

With a decoding delay budget uniformly set at 20ms for the majority of traffic types~\cite{3gpp}—Figure~\ref{fig:heatmap} illustrates heat maps that reveal how throughput and latency shift in response to variable resource allocation. The evidence from these maps decisively indicates \Name{}'s superior capacity to improve resource efficiency and elasticity at the network edge compared to Nuberu and BaseLine. \Name{} is able to provide enough decoding capacity (white region in Figure~\ref{fig:heatmap}) in various resource combination. Both Nuberu and BaseLine are constrained by their inability to offload DU workloads to remote servers, tethering their throughput capabilities exclusively to the CPU cores at hand. BaseLine, in particular, encounters an early throughput cliff (90\% decrease) as CPU resources dwindle (10 cores), with an increasing number of packets failing to meet the HARQ deadline and subsequently being discarded as irrecoverable decoding errors. Nuberu exhibits a marginal improvement due to its predictive handling of packets still undergoing decoding, which permits these packets to await idle CPU slots within the allotted 20ms delay budget. However, when edge resources reach their saturation point - evident in the darker areas of Figure~\ref{fig:throughput_heatmap}—the minimal HARQ deadline misses achieved through Nuberu's predictive approach are insufficient to avert a decline in throughput (90\% decline with 5 cores), as available slots within the 20 ms decoding window become increasingly elusive.
In contrast, \Name{} distinguishes itself by not only realizing a higher decoding throughput that decreases incrementally as edge resources are depleted, but also by leveraging MH bandwidth to sustain considerable decoding throughput levels, even when edge computing resources are substantially limited. Notably, \Name{} maintains up to 75\% of the maximal throughput capacity by offloading DU workloads through the MH network bandwidth and eliminating early decoding load through link adaptation.

Figure~\ref{fig:latency_heatma} shows that increasing CPU cores improves latency, while offloading does not directly affect latency. Although \Name{} incurs a modest latency increase due to edge buffering, it successfully maintains a 99th percentile latency of 20ms, comparable to BaseLine and Nuberu. This consistency demonstrates the effectiveness of the Earliest Deadline First (EDF) scheduling employed by \Name{}.
Figure~\ref{fig:throughput-latency} illustrates the relationship between throughput and latency as a function of the number of edge CPU cores under fully provisioned MH network conditions. \Name{} outperforms Nuberu with higher throughput and lower latency in all conditions. We also compare \Name{} without offloading to Nuberu, finding that in such cases, Nuberu performs similarly to \Name{}

In summary, \Name{}' strategy of decoding splitting and offloading optimizes network resource utilization and ensures robust performance across varying operational demands and resource combinations, establishing it as a crucial innovation to enhance 5G RAN efficiency.

\subsection{Evaluating the Scheduling Framework of \Name{}}
The scheduling framework of \Name{} is put to the test through a series of benchmarks aimed at assessing its performance under various workloads.

\begin{figure*}
    \centering
    \begin{subfigure}[b]{0.45\textwidth} 
        \centering
        \includegraphics[width=\linewidth]{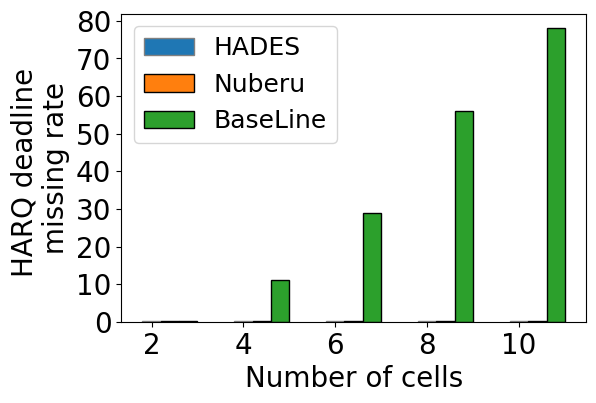}
        \caption{}
        \label{fig:edge_dead}
    \end{subfigure}
    \hfill
    \begin{subfigure}[b]{0.45\textwidth} 
        \centering
        \includegraphics[width=\linewidth]{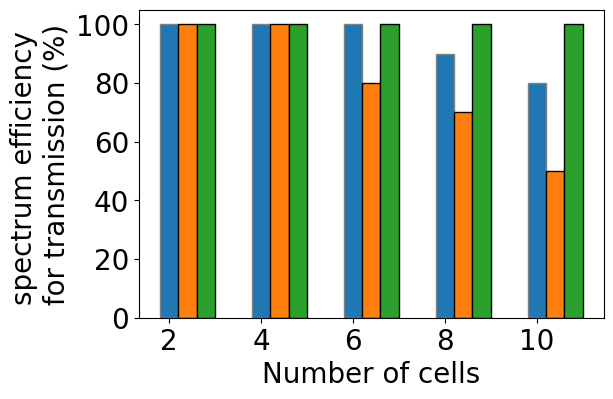}
        \caption{}
        \label{fig:spt}
    \end{subfigure}
    \hfill
    \begin{subfigure}[b]{0.45\textwidth} 
        \centering
        \includegraphics[width=\linewidth]{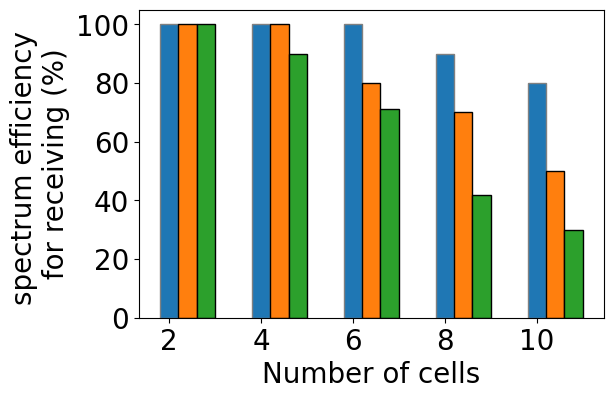}
        \caption{}
        \label{fig:spr}
    \end{subfigure}
    \hfill
    \begin{subfigure}[b]{0.45\textwidth} 
        \centering
        \includegraphics[width=\linewidth]{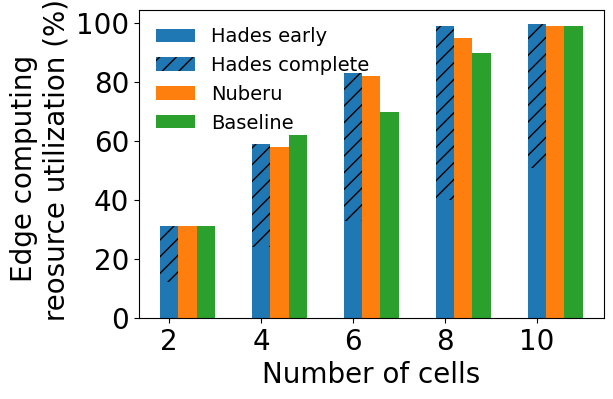}
        \caption{}
        \label{fig:edge_utlization}
    \end{subfigure}
    \caption{A comprehensive assessment across various cell configurations, showcasing (a) the edge HARQ deadline missing rate, (b) spectrum efficiency for transmission, (c) spectrum efficiency for receiving, and (d) overall edge resource utilization. The bar graphs compare \Name{} with Nuberu and Baseline, indicating \Name{}'s ability to minimize deadline misses and optimize resource use while maintaining high transmission and receiving efficiency.}
    \label{fig:schedulingEvaluation}
\end{figure*}

\textbf{Deadline Missing Rate Control}
The ability of \Name{} to manage HARQ feedback deadlines is a critical aspect of its performance. Figure~\ref{fig:edge_dead} demonstrates that \Name{} consistently maintains a low deadline missing rate, significantly below the 0.1\% configuration level, and well within acceptable bounds compared to the 5\% target retransmission rate prescribed by standards~\cite{3gpp}. In contrast, BaseLine suffers from a high rate of missed deadlines due to resource scarcity and lack of adaptive rate control, resulting in increased NACKs and retransmissions. Nuberu shows similar improvement in managing missed deadlines and BLER, although its PRB-based load adaptation tends to overcompensate as we show next.


\textbf{Impact on Spectrum Usage}
\coloredtext{red}{
The deadline missing rate directly affects how load is adapted through spectrum usage based on decoding capacities. Figures~\ref{fig:spt} and~\ref{fig:spr} quantify the impact of load adaptations on spectrum usage. Spectrum efficiency for transmission, shown in Figure~\ref{fig:spt}, represents bit rates for transmission and is influenced by the load adapter. Without load adaptation, the full bit rate is enabled, as seen in the baseline. Nuberu starts its load adaptation by limiting spectrum allocation at 6 cells, while \Name{} triggers load adaptation at 8 cells. 
Spectrum efficiency for receiving, depicted in Figure~\ref{fig:spr}, represents the ratio of bits successfully decoded. Overloading is indicated when reception is lower than transmission, a scenario in which BaseLine experiences a significant decline in reception due to overloading, which is also confirmed by the significant BLER in Figure~\ref{fig:edge_dead}. \Name{} strategically lowers spectrum efficiency to reduce the load on edge decoding, contrasting with Nuberu's PRB-based adaptation, which restricts spectrum allocation and can lead to higher underutilization and inefficiency.
}

\textbf{Maximizing Edge Resource Utilization}
Figure~\ref{fig:edge_utlization} illustrates \Name{}' strategic utilization of edge resources. \Name{} effectively uses periods of inactivity at the edge to facilitate completion decoding, especially after offloading tasks to remote servers via MH networks. This adaptive utilization contrasts with BaseLine, which shows lower resource usage due to its rigid adherence to strict decoding delay budgets. Like \Name{}, Nuberu leverages extended decoding delays post-prediction to enhance resource utilization. \Name{} further distinguishes itself by efficiently prioritizing prediction decoding when necessary, achieving superior resource utilization rates and offloading excess completion decoding to remote servers.
There are no differences in edge utilization when the systems are underloaded(2/4-cells) or excessively overloaded(10-cells). When underloaded, \Name{} and Nuberu perform similarly to BaseLine, with no need for load adaptation or offloading.

Overall, \Name{}’s scheduling framework demonstrates robust performance, adeptly managing deadlines, optimizing spectrum usage, and maximizing edge resource utilization even under constraints. \Name{} proves to be a sophisticated solution capable of intelligently navigating network load and resource allocation challenges, ensuring efficient operation across varying conditions.

\subsection{Traffic-Dependent Throughput and Efficiency}
The system throughput and efficiency of edge computing in uplink traffic decoding is highly influenced by the characteristics of the traffic. This section identifies the key factors that affect system performance in the context of uplink traffic.

\begin{figure}
\begin{subfigure}{\textwidth}
  \centering
  \includegraphics[width=\linewidth]{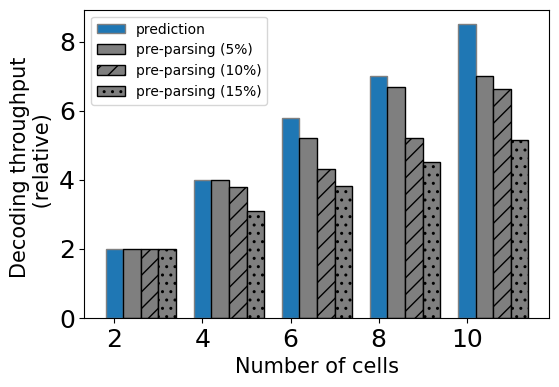}
  \caption{Throughput with different \\ pre-parsing ratios}
  \label{fig:throughput_overhead}
\end{subfigure}%
\hfill
\begin{subfigure}{\textwidth}
  \centering
  \includegraphics[width=\linewidth]{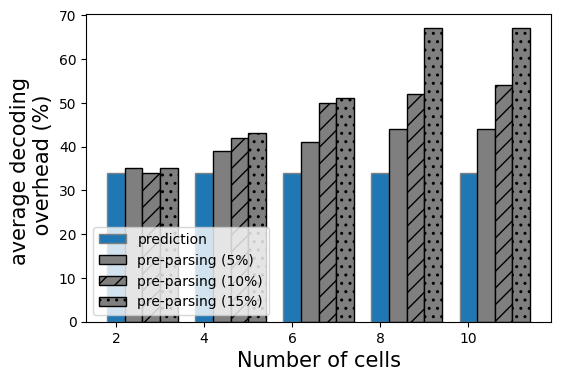}
  \caption{Iteration overhead with different \\ pre-parsing ratios}
  \label{fig:early_overhead}
\end{subfigure}
\caption{Impact of Pre-parsing Ratios on Decoding Efficiency: (a) Shows the throughput achieved with varying ratios of pre-parsing, illustrating how the increase in pre-parsing content (from 5\% to 15\%) affects the relative throughput across different numbers of cells. (b) Depicts the average decoding overhead associated with prediction and different levels of pre-parsing,}
\label{fig:overhead}
\end{figure}

\textbf{Early Decoding Overhead}
When the system is constrained in edge decoding capacity, the overhead load of early decoding directly affects system throughput. \Name{}' early decoding phase, encompassing prediction and pre-parsing, significantly impacts edge decoding efficiency. The workload of pre-parsing is influenced by the proportion of subheader bits within the overall PDU. Smaller packets with a larger share of subheader bits require more processing for pre-parsing.
Figure~\ref{fig:throughput_overhead} shows how varying pre-parsing bit ratios affect decoding throughput under different load conditions (2-10 cells traffic load). When the system is lightly loaded (2 cells), throughput is not significantly affected by pre-parsing load due to sufficient edge decoding capacity. However, as the system becomes more loaded, \Name{} experiences a throughput decline with an increasing ratio of pre-parsing bits. This decline is more significant under heavier loads, decrease by 30\% with 15\% pre-parsing bits when we have 10cells load.

Figure~\ref{fig:early_overhead} further illustrates the load implications across different traffic scenarios, highlighting that a higher pre-parsing load correlates with a decline in throughput. This is because more workload needs to be processed under low latency constraints, exacerbating the impact of limited edge decoding capacity.


\begin{figure}
    \centering
        \begin{subfigure}[t]{\textwidth} 
            \centering
            \includegraphics[width=\linewidth]{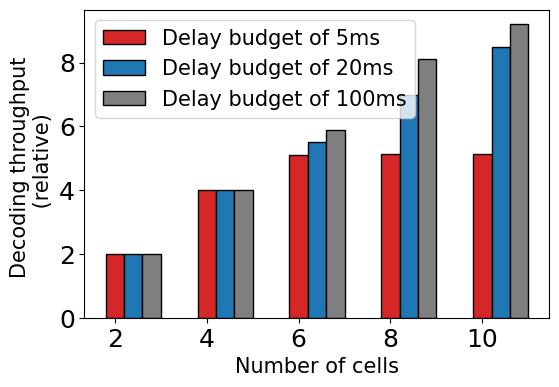}
            \caption{MH latency 10ms}
            \label{fig:throughput_delaybudget10}
        \end{subfigure}%
        \hfill
        \begin{subfigure}[t]{\textwidth} 
            \centering
            \includegraphics[width=\linewidth]{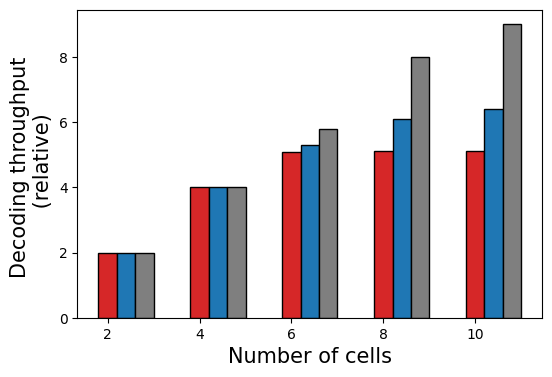}
            \caption{\coloredtext{red}{MH latency 15ms}}
            \label{fig:throughput_delaybudget15}
        \end{subfigure}
        \caption{The relative decoding throughput achieved for different delay budgets (5ms, 20ms, 100ms) across an increasing number of cells, indicating how latency constraints affect throughput capacity.}
        \label{fig:throughput_delaybudget}
    \hfill
    \begin{subfigure}[t]{\textwidth} 
        \centering
        \includegraphics[width=\linewidth]{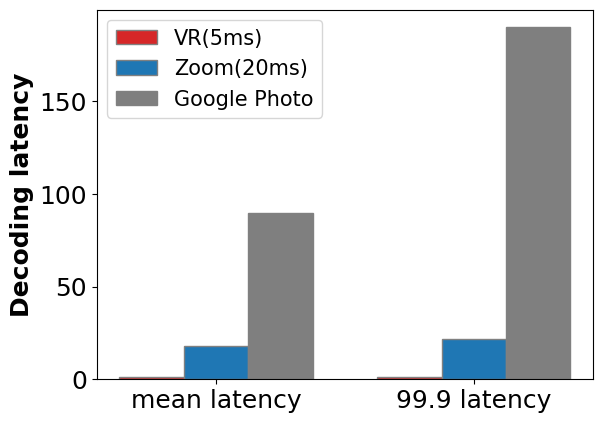}
        \caption{The decoding latency experienced by different types of traffic.}
        \label{fig:mixeddelay}
    \end{subfigure}
\end{figure}

\textbf{Impact of Delay Budgets on Throughput}
The delay budgets assigned to uplink traffic critically determine the potential for offloading and leveraging idle edge CPU cycles. Figure~\ref{fig:throughput_delaybudget} presents the relative throughput achievable with six CPU cores under varied decoding delay budgets. Larger delay budgets enable more efficient utilization of edge CPU resources by allowing longer waits for available CPU slots, leading to significant throughput improvements, especially when comparing $20ms$ to $5ms$ delay budgets. However, benefits diminish with delay budgets extending beyond $20ms$, as opportunities to offload to remote servers decrease. There is an almost doubling of throughput
from $5ms$ traffic to $20ms$ traffic.
\coloredtext{red}{
Alternatively, large MH latency deters the offloading.
As we increase the MH latency from $10ms$ (~\ref{fig:throughput_delaybudget10}) to $15ms$ (~\ref{fig:throughput_delaybudget15}), the throughput of traffic with $20ms$ delay budget declines significant during overloaded situation (8cells and 10cells).
}

In real-world scenarios, uplink traffic often includes a variety of services, each with unique latency requirements. \Name{} effectively handles this diversity through its pre-parsing feature, which categorizes and manages traffic based on their delay budgets, as shown in Figure~\ref{fig:mixeddelay}. 
Our evaluation simulated a mix of service categories
: 5\% of the traffic was allocated to VR applications with a stringent 5ms delay budget, 50\% to video live streaming services like Zoom with a 20ms delay budget, and the remaining to best-effort traffic like Google Photos with a more lenient 100ms delay budget. 
The traffic characteristics, including data size and rate, were modeled on real application traces, leading to scenarios where a single MAC PDU packet might carry data from multiple service types.
Figure~\ref{fig:mixeddelay} demonstrates that both the mean latency and 99.9 percentile latency are differentiated for the three different traffic types and perform under the QoS requirements. Although our test distribution might not precisely reflect typical real-world traffic patterns, the results provide valuable insights. They demonstrate \Name{}’s capacity for adaptive traffic management, showcasing its ability to dynamically and efficiently fulfill the varying delay requirements of mixed traffic.


%% file: conclusion.tex
\section{Conclusion}

In this study, we have presented \Name{}, a novel hierarchical adaptive decoding system designed to enhance the efficiency and elasticity of edge computing for uplink decoding in vRANs. \Name{} capitalizes on the hierarchical nature of vRAN architecture, optimizing the use of MH bandwidth, and addressing the challenges posed by vRAN edge computing capacity constraints.

By innovatively splitting uplink decoding tasks into early and completion decoding phases, \Name{} adapts to the dynamic environment of edge computing and MH bandwidth resources. Our system’s flexibility in scheduling and offloading decisions has demonstrated significant improvements in computational efficiency and latency management. Through comprehensive evaluation, we have shown that \Name{} can maintain high decoding throughput and manage latency effectively even under stringent edge resource limitations.

The key insights from our evaluation indicate that \Name{} has virtually no decoding capacity decrease
with up to half of the edge decoding capacity reduction and \Name{} can scale up to 75\% of the maximum throughput capacity through effective offloading strategies. This is particularly notable when compared to existing systems like Nuberu and BaseLine, which do not support workload migration and thus, are bound by the available CPU cores at the edge. Furthermore, \Name{} proves its robustness in handling mixed traffic scenarios with varying delay budgets, showcasing its ability to prioritize tasks based on the stringent latency requirements of 5G networks.
